\documentclass[a4paper,11pt]{article}
\usepackage{yfonts}
\usepackage{grffile}
\usepackage{ifthen} 
\usepackage{graphicx}
\graphicspath{ {./images/}} 
\usepackage{adjustbox}
\usepackage{amsmath}
\usepackage{amssymb}
\usepackage{verbatim}


\title{String Invention, Viable 3-3-1 Model, Dark Matter Black Holes}

\author{{Holger B. Nielsen}\\
{Niels Bohr Institute, {2200} Copenhagen, Denmark; hbech@nbi.ku.dk}}

\begin{document}
\maketitle
\begin{abstract}
  With our very limited memories, we provide a brief review of Paul Frampton's
    memories of the discovery of the Veneziano model, with this indeed being
    string theory, with Y. Nambu,
 and, secondly, his
 3-3-1 theory. 
 The latter is, indeed, a non-excluded replacement for the
 Standard Model with triangle anomalies being cancelled, as they must in a
 truly
 viable theory. 
  It even needs (essentially) three as the family number! Moreover,
 primordial black holes as dark matter is mentioned. We end with a review of
 my own very speculative, utterly recent idea that for the purpose of the
 classical approximation, we could, using the functional integral as our
 rudimentary assumption taken over from quantum mechanics, obtain the equations
 of motion without the, in our opinion, very mysterious imaginary unit $i$,
 which usually occurs as a factor in the exponent of the functional
 integrand, which is this $i$ times the action. The functional integral
 without the mysterious $i$ leads to the prediction of some of the strongest
 features in cosmology, and also seems to argue for as few
 black holes as possible
 and for the cosmological constant being zero.
\end{abstract}


\section{Introduction}

Thinking about Paul, first, I do remember our nice time together in CERN
when we were younger, getting nice dinners and visiting the mountains,
once even with a couple of diplomat girls. However, first of all, we were working,
and Paul was, in addition to working with me, etc.,\ writing his book~\cite{book} on
dual models
with its several chapters. I had met Paul earlier than at CERN, I think in DESY.
Additionally, once he even met my mother and uncle in Copenhagen, and I remember
we were at a place in the far end of Nyhavn, and that was rather special
in the way that I think my mother and uncle met exceedingly few of my
colleagues. 

In Geneva, we went to a lot of good restaurants together and there were
many possibilities. During the same stay in CERN, I also worked with Lars Brink, and also, Colin Froggatt,
with whom I am still working a lot, was there. In addition, Paul and I worked a bit together much later, and I have not given up the
hope of us getting even more work together; in fact, I am going to
Corfu soon, and Paul is to be there too. 

\subsection*{Plan of Paper}

In the next section, I review the very early stages
of finding the string from the Veneziano model as conducted by Frampton and Nambu. In Section \ref{sec3}, we review the 3-3-1 model, which Paul created and
which has been studied a lot, because it is a possible replacement for the
Standard Model, which really could be true still. In Section \ref{beyond},
we look at which arguments one would speculate Nature may have used for
finally selecting the model, 3-3-1 or the Standard Model.
In Section \ref{BH}, we move on to black holes, mainly with the idea that they should
be dark matter as this is Paul's favorite dark matter model. Then, in Section
\ref{own}, I refer to my own work, which attempts as a 
Random Dynamics project
to derive quantum mechanics but instead discusses how one makes a classical
theory from functional integral formulation with the, to me, mysterious
{\bf {imaginary unit} $i$}, in a fundamental theory such as quantum mechanics,
{\bf {removed}}. 
Section \ref{conclusion} provides a conclusion and birthday wishes.

\section{Very Early String Theory}
\label{String}
In the celebration of Nambu~\cite{Nambu}, Paul reveals how I,~\cite{stringbirthHBN}, and
Lenny Susskind~\cite{stringbirthLS} should feel very lucky that 
Nambu~\cite{stringbirthYN, stringbirthYN2}
was a bit slow in publishing 
about the string from factorization of the dual model. (See also the book
on the Birth of String Theory~\cite{stringbirthbook}).

Indeed, after a question by Nambu, Paul worked on factorizing the factor
\linebreak
\mbox{$(1-x)^{-2\alpha' p_2\cdot p_3}$} in the Veneziano model to what would now be
called string variables. The question was:

Can the $t$-dependence be factorized {as} 

\begin{eqnarray}
(1 - x)^{-2\alpha' p_2\cdot p_3} = F (p_2)G(p_3) ?
 \end{eqnarray}
where this is the factor in the Veneziano model notation
\begin{eqnarray}
 A(s,t) &=& \int_0^1x^{-\alpha(s)-1}(1-x)^{-\alpha(t)-1}dx.
 \end{eqnarray}

Paul carried out the factorization by writing
\begin{eqnarray}
 (1-x) &=& \exp(\ln(1-x)),
 \end{eqnarray}
and expanding
\begin{eqnarray}
 \ln(1-x) &=& -\sum \frac{x^n}{n}.
\end{eqnarray}

He reached the factorization
\begin{eqnarray}
F (p)& =& exp
\left (
i\sqrt{2}\alpha' p_{\mu}
\sum_1^{\infty}\frac{a_{\mu}^{(n)}x^n}{\sqrt{n}}\right )\\ 
G(p)& =&\ exp
\left (
i\sqrt{2}\alpha' p_{\mu}
\sum_1^{\infty}\frac{a_{\mu}^{(n)\dagger}}{\sqrt{n}}\right )
\end{eqnarray}

It is easy to check that the explicit solution is
\begin{eqnarray}
 (1-x)^{-2\alpha'p_3\cdot p_2}&=& <0|F(p_2)G(p_3)|0>.
 \end{eqnarray}

This was all very early, even compared to my own also unpublished version
---even the first version, 
which was the same but without the word ``almost'' in the title. 
 In fact, Knud Hansen, an experimentalist at the Niels Bohr Institute, commented on my first title without the ``almost'' so that I put in this word and then produced the title ``An almost physical Interpretation of Dual
Model''~\cite{stringbirthHBN}.

\section{Extension of Standard Model That Could Work}
\label{sec3}

The so-called 3-3-1 model\cite{Pl,PRL69}, of which one can have some
slightly different
variations (see e.g.,~\cite{DongSi}), is a model carefully worked out to
have {\bf {no anomalies}} from the
triangle diagrams for the fermions in the model. This
model~\cite{PRL69} is described
 as $SU(3)_c\times SU(3)_L\times U(1)_N$, or thinking with O`Raifeartaigh
~\cite{ORaifeartaigh} on gauge {\bf {groups}} rather than just the Lie
 algebra, $U(3)_c\times SU(3)_L$, 
 and the leptons have their left pair put together with the antiparticle of the
 right-handed singlet of the
 usual Standard Model, into a single antitriplet representation
 for the $SU(3)_L$ representation
 \begin{eqnarray}
 \psi_{aL} &=& \left ( \begin{matrix}
 \nu_a\\
 l_a'\\
 l_a^{'c}
 \end{matrix}\right ) \sim ({\bf 3}, 0)
 \end{eqnarray}

\begin{figure}
   \includegraphics[scale=2]{
     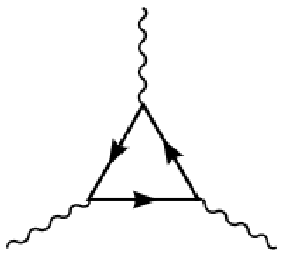}
 \caption{\label{fig1} {In this} {diagram,} the three outgoing
 gauge boson (propagator) attachments denoting the gauge bosons
 from the gauge group $SU(3)_L$ replaced the weak $SU(2)$ and
 some of the $U(1)$ in the Standard Model, which is discussed in the text.}
 \end{figure}

 For the quarks, we do not in the Standard Model have the $SU(2)$ singlet
 as for the leptons, and thus, instead, a new particle $J_a$ is introduced to
 complete the triplets under the $SU(3)_L$. The naive attempt would be to
 let all the three families of quarks we know also be represented as the
 lepton $SU(2)$-doublets as, e.g., the first family
 \begin{eqnarray}
 Q_{1L}&=& \left ( \begin{matrix}u_1'\\
 d_1'\\
 J_1
 \end{matrix}\right )
 \sim \left({\bf 3},2/3\right ),
 \end{eqnarray}
 but if we had all the quark families represented this way, then the
 anomaly from a triangle diagram (see Figure~\ref{fig1}) with three external
 gauge particles from
 group $SU(3)_L$ would add up and the model would not be anomaly free.

 Instead, it must now be repaired by representing instead one of the
 quark families by the representation
 \begin{eqnarray}
 Q_{\alpha L} &=& \left( \begin{matrix} J_{\alpha}'\\
 u_{\alpha}'\\
 d_{\alpha}'
 \end{matrix}\right )\sim \left({\bf 3*},-1/3\right )
 \hbox{ for $\alpha$ = 2, 3}.
 \end{eqnarray}
 
 Then, we can easily see that we can achieve a cancellation
 of the anomalies from the triangle diagram with three outgoing gauge
 particles from the $SU(3)_L$ group, with one extending the weak $SU(2)$
 and, partly, the $U(1)$ of the Standard Model. We just have to know
 that the diagram anomaly is non-zero because it is obvious that the anomaly
 contribution is simply proportional to the number of left-handed fermions
 going around the triangle in the diagram. Thus, we must have arranged that
 there are equally as many left-handed particles going around
 with the representation ${\bf 3}$ as with the conjugate
 ${\bf 3*}$.

 Since whatever choice we make with the different quark families,
 we always
 obtain them in triples because of the color representation, so they will
 always also
 participate with a multiple of three of the triangle diagram for
 three external $SU(3)_L$ gauge bosons, unless we choose higher than the
 sextet representations
 for them. Thus, under this attempt to keep close to the Standard Model and
 at least not postulate, e.g., $SU(3)_c$-octet-colored fermions, there must be
 lepton families in a multiple of three. With the just-sketched trick
 of having two quark families in the ${\bf 3*}$, while there is only one in the
 ${\bf 3}$, we can indeed obtain the no anomaly conditions satisfied by the
 relations
 \begin{eqnarray}
 (\hbox{``1 family''}*3 + 3 \hbox{ leptons}) anomaly(\hbox{{\bf 3}
 $SU(3)_L$'s})
 & + &\nonumber\\
 \hbox{``2 families''}*3 *anomaly(\hbox{{\bf 3*} of $SU(3)_L$'s})
 &\propto& \nonumber\\
 3+3-2*3&=&0.
 \end{eqnarray}
 
 Models of this type were made without any anomalies, so that they really
 could be true, and so close to the Standard Model, that they cannot be
 excluded yet. It is remarkable that it has its predictions much
 closer to the experiments than say grand unification theories, which
 only provide a number of coincidences of fine structure constants and
 proton decay,
 which are very remote in energy scale, while Paul's 3-3-1 model(s)
 have new particles much closer. (Actually there are limits to how far
 they can be put away). One test of 3-3-1 is the bilepton resonance in same sign leptons.

 For the good reason of being, in this sense, closer to possible reality,
 it has already been well studied and deserved lots
 of citations; actually, if it really were true, it could be found very soon
 that it was indeed the case. 

 There are of course obvious dangers for the 3-3-1 model, namely that the
 relations
 between measured quantities, which come from the Standard Model as successfully predicted,
 do not quite come out in the 3-3-1 models. The question of whether the 3-3-1
 model can manage to organize the successful predictions such as no flavor
 changing neutral currents (i.e. bounds for flavor changing neutral currents
 FCNC) and the $\rho$ parameter
 \begin{eqnarray}
   \rho &=& \frac{m_W^2}{m_Z^2\cos^2\theta_W} 
   \end{eqnarray}
 is complicated because the model needs some scalar triplets under the
 $SU(3)_L$ in order to break the symmetry down to the Standard Model.
 In fact, one has three scalars that are triplets under the $SU(3)_L$
 (the one connected to weak interactions), and one can make different models
 by leaving out one or the other of these scalars.
 It turns out, however, that there is a need to fit all the scales of the
 scalar triplets to adjust the 3-3-1 model to match the known
 conditions~\cite{DongSi}. Thus, some simplified version(s) are excluded.

 \section{Some Day We Shall Explain Whether the 3-3-1 or the Old SM Is Right}
\label{beyond}
 The great thing is that the 3-3-1 model is closer than many other theories
 proposed, and then it should not take an unrealistically long time
 to find out if it is right or if the Standard Model is the right one.
 At that time,
 one would likely give some principles by word that point to the right model, in order to find out what the principles are
 behind the even deeper physics, which determines which model we should find
 in the LHC energy range. 
  One might set out with some questions like:
 \begin{itemize}
 \item{{\bf {Is the number of families determined?}}} 
 The 3-3-1 model has the feature, via the disappearance of anomalies,
 needing (a multiple of) three families. 
  Insisting on the asymptotic freedom of QCD requires exactly three families.

 This is of course a great victory for the 3-3-1 model, which, thus, does
 predict the number of families, and thus, it should support our belief
 in the model. However, is this something we should think about for the next level
 of theories so as to bring us onto the track of the next
 level of theories? 
  Presumably, the next more fundamental
 level theory would not care if there some understanding was left
 of the number of families, or we would be left with that for the next
 level.

 \item{{\bf {Small representations of the fermions}}}
 This is seemingly a good principle to apply for the details of both
 the 3-3-1 model and the Standard Model, since both (types of) models
 have the smallest non-trivial representations of the groups at hand.

 One would have to make the concept of smallness of
 representations very detailed to make assuming such a detailed definition of
 smallness a good argument for why Nature should choose the one or 
  the
 other (from a deeper physics point of view).

 \item{{\bf {The gauge group}}} Philosophizing over the gauge group, you might
 propose thoughts like this:

 If in a deeper physics there are principles or mechanisms that
 favor making the gauge group at lower energies in some special simple group,
 then you might expect it to be better to use the same group several
 times and make the full group or Lie algebra a cross-product of several
 isomorphic groups, in the same way the 3-3-1 model has $SU(3)$ used twice.
 This saying would of course give a speculative argument in favor
 of the 3-3-1 model (relative to the Standard Model).

 Thus with such thinking in mind you would, if the Standard Model
 should finally turn out to be the right one, wonder very much why a gauge
 group was chosen with two different simple groups in it, in addition
 to the
 $U(1)$. I would say this mystery might have a possible explanation
 in the next item.
 \item{{\bf {Which group has the strongest connection between the Lie algebras by
 rule(s) of allowed representation combinations?}}}
 I would say that it looks like Nature has chosen the Standard Model
 partly because it loves that the gauge {\bf {group}} has obtained the
 center of the covering group divided out by a discrete group so
 as to connect the different Lie algebra cross-product factors. 
 Having gauge {\bf {groups}} is the O'Raifeartaigh way of looking at
~\cite{ORaifeartaigh} the quantization rules connecting
 the allowed combinations of (irreducible) representations of the
 different Lie algebras.

 In fact, the Standard Model has a quantization rule for the
 $U(1)$ charge usually called y/2, which relates it to {\em both}
 the representation of the color $SU(3)_c$
 and the weak \mbox{$SU(2)$ requiring}
 \begin{eqnarray}
 \hbox{``triality'' for $SU(3)_c$} +s_W +y/2&=& 0 \; (mod\; 1),
 \label{con}
 \end{eqnarray}
 which is of course the rule that ensures the quantization of the electric
 charge as it is believed in the Standard Model. Here,
 $s_W$ is the representation classifying number for the Standard
 Model weak $SU(2)$: i.e., the ($2s_W+1$)-plet. 

 You could not connect the given Lie algebras more by such a
 quantization rule than by this (\ref{con}).

 In the 3-3-1 model, there is only such a quantization rule connection
 between the $U(1)_X$ and the color $SU(3)_c$. The $SU(3)_L$ is not
 connected this way.
 Thus, the 3-3-1 model is not quite as strongly connected as the
 Standard Model, which has such a quantization rule connection between
 all three 
 Lie algebra cross-product factors \textgoth{u}(1), \textswab{s:u}(2),
 and
 \textswab{s:u}(3). In fact, the most intertangling quantization
 rule for the $U(1)$ charge $y/2$ is postulated for the Standard Model.
 Now, such a high degree of complication for the quantization rule
 can only be made provided the orders of the centers of the
 simple non-abelian groups, in the Standard Model,
 $SU(2)$ and $SU(3)$ are {\bf {incommensurable/mutually prime.}}
 The two smallest incommensurable natural numbers that can be
 used in $SU(N)$-groups are actually 2 and 3.

 For example, if you have the Lie algebras as in 3-3-1, \textgoth{{u}}(1), \textgoth{{s:u}}(3), \textgoth{{s:u}}(3), you cannot make a non-trivial
 quantization rule connecting the \textgoth{{u}}(1) to {\em both}
 \textgoth{{s:u}}(3)'s. (3 and 3 are, namely, not incommensurable).

 Thus, one could claim that the Standard Model would have to have the
 two different Lie algebras in order that one could then make
 such quantization rules for {\em both} non-abelian algebras.

 \item{{\bf {Skewness}}} Once, I and Niels Brene~\cite{skewness1,skewness2,skewness3}, in our search
 for some characteristic properties of the Standard Model, proposed a
 concept of skewness that should basically mean that the
 (gauge) {\bf {group}}
 (and here we really thought of the group in O'Raifeartaigh's sense~\cite{ORaifeartaigh} rather than just the Lie
 algebra) should have very few (outer) automorphisms compared to the
 rank. 
  We found that the appropriately defined Standard model {\bf
 {group}} $S(U(3)\times U(2))$ (which is defined as the group of
 $5\times 5$ matrices composed from the $SU(2)$ and the $SU(3)$
 matrices and imposing the condition given by the $S$, that the
 determinant of the $5 \times 5$ matrix should be 1). would be pointed
 out as {\bf {most skew}}.

 The 3-3-1 model group is $U(3)_c\times SU(3)_L$ because
 there is a rule connecting the $U(1)_X$ charges so that they have
 1/3 modulus 1 for triplets of the color $SU(3)$, while
 the ``$U(1)_X$ charge'' is not connected similarly to the
 $SU(3)_L$ representation. 

 This {\bf {group}} of the 3-3-1 model at least has the interesting
 sign of skewness, in that
 just the colored $SU(3)$ obtained a rule connecting it to the $U(1)_X$,
 while the other $SU(3)$ has no such connection to the $U(1)_X$.
 This, namely means that the obvious outer automorphism of the
 group $U(3)_c\times SU(3)_L$ consisting of permuting the
 two $SU(3)$-subgroups is prevented, so that at least this
 automorphism is not there.

 However, you can make a complex conjugation of the $U(3)_c$ and of the
 $SU(3)_L$ separately, and thus, the outer automorphism group
 becomes ${\bf Z}_2 \times {\bf Z}_2$, while the Standard Model group
 only has an outer automorphism group ${\bf Z}_2$.
 \item{{\bf {Holy number 3}}} The 3-3-1 model has ``the holy number 3''
 we could fantasize first in the two SU(3) groups, and then from there,
 it is transferred with the anomaly avoidance to the number of families.
 Thus, it is really characterized by a ``holy number''.

 I do not see that you could say the same thing about the Standard
 Model.
 \item{\bf {Unification}}
 In seeking theoretical stories that could be used to
 say a posteriori why Nature should have chosen one model or the
 other, the possibility of putting the model into a grand unification
 model is almost a must to be mentioned. I am afraid I shall not be able
 to study the possibilities of making a grand unification model
 extension of the 3-3-1 model before Paul is more than 80 years old,
 but we can hope to find out such possibilities when he gets to 90.
 It is not fitting in at all that $SU(5)$ is the
 starting point for most unifications for the Standard Model. 

 It is, however, well known that the success for unification
 of the Standard Model without some helping complications
 like super-symmetry has not been so great again~\cite{SU5}.

 I have myself followed the attempt by Norma Mankoc Borstnik~\cite{Norma}
 and collaborators of having 10 extra dimensions in addition to the
 4, which we see clearly, and $SO(10)$ is one of the studied unification
 extensions of the Standard Model.

 However, very recently I have been keen on accepting that the
 $SU(5)$ symmetry could be only approximate~\cite{Approximate} and
 further gauge particles
 in it compared to the Standard Model should not exist in reality.
 Really, it is a lattice gauge theory I propose in which $SU(5)$
 symmetry comes in the classical approximation but is broken
 by quantum corrections; you just have to multiply the quantum
 correction by a factor of 3 obtained by giving each family its own
 lattice.
 \end{itemize}
\section{Black Holes}
\label{BH}

For the purpose of the present article, I thought it best to read a
little bit of Paul's papers, preferably about, e.g., dark matter.

Of course, I come in as one who definitely does not believe in dark matter
being primordial black holes, in as far I have written with Colin Froggatt, who also was in CERN at the same time Paul and I were there, a long series
of works about it being small (it could at first come from much
bigger regions of an alternative vacuum, then contract, or still
survive today)
macroscopic objects; but then Paul's black holes sound very convincing!

After all, the black hole theory of dark matter is genuinely
without new physics, whereas mine and Colin Froggatt's model
\cite{Dark1,Dark2,Tunguska,supernova,Corfu2017,Corfu2019,theline,Bled20,
  Bled21,Higgs} only
formally is without new physics because it needs two phases of vacuum.
Such two vacuum phases would a priori need new physics, unless we have
such remarkably good luck that the calculations in QCD of the properties of vacuum for
different values of the quark masses~\cite{Columbia1} should reveal a non-trivial behavior, and
the experimental quark mass combination should just lie on the phase transition.
At least one of the authors~\cite{oa} looking for phase transitions represented
the chance that we could have hope of really having no need for new
physics in our model, by admitting with question marks on his phase plot
that he did not know if the experimental quark combination was in the one phase or
the other. Thus, it could miraculously be on the very border line.
If so, there could be two phases existing in extensive regions of spacetime
in the universe, e.g., inside and outside dark matter pearls~\cite{Dark1}.

Well, it must be admitted, that even if there was the appropriate phase
transition between different phases that would still need the milder amount
of new physics, there should be some principle, some law of nature,
ensuring that the quark masses, say, would have just the right masses to be
just on the phase border, so that more than one phase could be realized with
essentially the same energy density. This hypothesis we have
talked much about under the name ``Multiple Point Critically Principle''
\cite{Kannike, Selke, Don, Higgs, MPPDC}
(MPP), and we once had the luck of predicting~\cite{Higgs} the Higgs mass from assuming
this hypothesis, before the Higgs was produced at LHC!

Actually, it turned out that what we thought was the main hypothesis in our
MPP, namely that
different phases of vacuum should have the same energy density or cosmological
constant, Dvali and earlier Zeldovic had already proposed as a
theorem~\cite{Dvali, Zeldovic, Okun}. 
 Doubts concerning this theorem are
discussed by C. Gross, Strumia, et al.~\cite{CGross}.


\subsection*{PBH Is a Possibility for Rather Heavy Dark Matter Particles}

Believing in Hawking radiation, the primordial black holes (PBHs) lighter than
$10^{15}g=10^{12}$~kg would have radiated away or would be just about doing it today.
This is a rather high mass compared to what is speculated in other models, and
it has, of course, consequences for how high a number of incidences with the Earth
we can have. We had, ourselves, a speculative model for dark matter once, in
which the Tunguska event in which the trees were thrown down or even up
by a big explosion in a 70 km large region should be due to the fall of
a dark matter particle. This Tunguska particle was from the
assumption that one fell on Earth every hundred years, and using the dark
matter density 0.3~GeV/cm$^2$ in the solar system, having a mass of about
$1.4*10^8$~kg. The PBHs have to be a few thousand times heavier, and thus,
correspondingly, more seldomly hitting the Earth.

If so, it is of course excluded that the PBH component of the dark matter
could be what is observed by the DAMA-LIBRA experiment~\cite{DAMA}.

Of course, this DAMA-LIBRA experiment~\cite{DAMA} is still rather in
contradiction
(one found in 2021, e.g., an article with the title
``Goodbye, DAMA/LIBRA: World’s Most Controversial Dark Matter Experiment
Fails Replication Test'' by
Ethan Siegel, alluding to the disagreeing Anais~\cite{Anais} experiment
with another underground search for dark matter, which actually find nothing
instead of confirming DAMA). 
Thus,
one might attempt to declare DAMA as seeing something else or being
mistaken somehow. If we shall uphold that DAMA saw many events of
dark matter (whatever), it cannot be primordial black holes, so we rather
must have several different components in the dark
matter if there are also the PBHs. DAMA-Libra has achieved C.L. for
the full exposure
(2.86 t~{$\times$}~yr) 13.7$\sigma$. However, the Anais experiment~\cite{Anais} sees no dark 
matter, i.e., a rate of $0.0003 \pm 0.0037$~cpd/kg/kev against DAMA-LIBRA of $0.0102\pm
0.0008$ cpd/kg/keV for a range 2keV to 6 keV. This rate means that DAMA-LIBRA sees
a season-varying amplitude of counts per day per kg detector per
bin of 1 keV energy in their scintillators of 0.0102 with uncertainty 8\%.
The Anais result is thus in contradiction with DAMA-LIBRA.

Froggatt and I propose that this contradiction comes about because DAMA-LIBRA is 1400 m down in the
earth, while the other experiments such as Anais are typically higher up towards the
earth's surface, so that the story could be made that the dark matter runs
fast through the near surface region and does not have time to radiate
electrons or X-rays before it gets stopped in the DAMA-LIBRA region.
Maxim Khlopov's model~\cite{Khlopov} has it that the dark matter of his 0-helium
should be stopped and make a nuclear interaction with nuclei inside the earth.
This could also mean that it first became active deeper down at the DAMA depth.
Such models have the chance of obtaining more counts for DAMA-LIBRA than for
the higher up experiments. Most of the underground experiments use xenon as the
scintillator liquid. The fact that it is a liquid could easily mean that
dark matter of a type that should be stopped or move slowly to be observed
would not be seen in the liquid xenon experiments. Alone, gravity might drive it
too fast through the detector if it is fluid. 

\section{My Own Very Recent Work on ``Kinetic Energy Being Unwanted''}
\label{own}

With the bad excuse that my recent work~\cite{BL} tends to predict that
black holes should
be kept only as a small fraction of the present energy density in the universe,
it should be preferably inline with the relativistic contributions from the photons and
the neutrinos, which have, respectively,
\begin{eqnarray}
 \Omega_{\gamma}h^2 &=& 2.480 \times 10^{-5}\\
 \hbox{and } \Omega_{\nu}h^2 &=& 1.68 \times 10^{-5},
\end{eqnarray}
namely, e.g., as~\cite{E} says, that about 0.04\% of the energy density
(=critical density) should be black holes. Already this is much compared to
what many would have thought, but of course, if the dark matter really consisted
of (primordial) black holes, then the $\Omega_{BH}$ would not be
the here-estimated $4\times10^{-4}$ but up in the $0.2$ region. My crazy idea
is in the series of ideas~\cite{Ninomiya1, Ninomiya2, Keiichi1, Keiichi2, Keiichi3,Influence, influenceColin,
 Don} in which I speculated about a theory that seeks to
combine physical equations of motion and initial states conditions, which
we in cosmology make assumptions about similarly
to how we make assumptions about the details of the laws of nature, the
Lagrangian density, and the system of particles that exist. Thus, such a
united theory should in principle tell, from the same formalism, about
how the Universe started and even how it shall end and about the laws of nature,
when you have put in the appropriate extra stuff, the Lagrangian, say. I long
worked
on that with Masao Ninomiya~\cite{Nino, Ninomiya1, Ninomiya2} and Keiichi
Nagao~\cite{Keiichi1, Keiichi2, Keiichi3, automatic, Formulation}, and we obtained the encouraging
result
that taking the Lagrangian or action to have complex coefficients in it
{\bf {would not be seen in the effective equations of motions resulting}}.
(The very earliest thoughts on this future influence might have been with
Colin Froggatt~\cite{influenceColin} and Don Bennett~\cite{Don}).
The only new information from such a complex action should be that it also makes sayings
about the initial state conscience. 
In the newest idea, which I talked about in the Bled Workshop on
``What comes beyond the Standard Models''~\cite{BL}, I want to say that
the $i=\sqrt{-1}$ in quantum mechanics really is a bit strange; should such a
fundamental theory as quantum mechanics really be based on the complex
numbers?

Should Nature really in the so fundamental quantum mechanics take up these
a priori just invented numbers, which Gerolamo Cardano around 1545 in his
Ars Magna~\cite{Ca} was inventing, although his understanding was rudimentary?
He even later described the complex numbers as being ```as subtle as they
are useless''

My recent work can be considered an attempt to ask: how needed really is this, is $i$ a priori not what you would expect Nature to choose? 

Let us {\bf {take over from quantum mechanics}} the idea of the functional
integral,
written almost symbolically
\begin{eqnarray}
 \hbox{Functional integral } &=& \int \exp(\frac{i}{\hbar}S[history])
 {\cal D} history.
 \end{eqnarray}

If you were thinking, as we have in mind here, on the development from the
beginning of time $t$ to the end of time, you might without knowing better
think that you should most simply obtain the expectation value for an operator
$O$ at some moment of time $t$, i.e., the expectation value of the ``Heisenberg
operator'' $O(t)$, by looking at the variable $O$ written in terms of the
variables
used in the functional integral taken at the time $t$ as constructed in
terms of these variables at $t$. That is to say that with the introduction also
of a normalization-denominator, one would naively propose to use as the
expectation value
\begin{eqnarray}
 <O(t)> &=& \frac{\int_{with\; |i>\; and \; |f>}O(t)\exp(\frac{i}{\hbar}
 S[history])
 {\cal D}history}{\int_{with\; |i>\; and \; |f>}\exp(\frac{i}{\hbar}
 S[history])
 {\cal D}history}, 
\end{eqnarray}
where I have also alluded to the fact that to make this expression meaningful, you
need at the earliest time to give a boundary condition state $|i>$ and also
one at the very final time $|f>$. This naive construction of the expectation
value $<O(t)>$ does in general {\bf {not}} give a really good expectation
value in as far as it is typically {\bf {not even a real number}}. Rather, this
functional integral naive attempt to construct an expectation value in a
seemingly not so bad way, is what is called {\bf {the weak value}}~\cite{wv} of the
operator $O$ at time $t$. 

The word $history$ is here used to denote the general path in the thinkable
development of the Universe from the beginning to the end, and $S$ is of course
the action functional. Really, you can see that this weak value is more like
a matrix element than a genuine expectation value, in as far as using the Heisenberg
picture, the weak value becomes
\begin{eqnarray}
<O(t)>_{weak\; value} &=& \frac{<f|O(t)|i>}{<f|i>} \hbox{ (Heisenberg picture)}.
\end{eqnarray}

We (Keiichi Nagao and I) made theorems about when it becomes
real~\cite{reality} for an
operator or dynamical variable $O$, which really means the Hermitian of
actually the slightly modified Hermiticity we required (``Q-hermiticity'').
There is one case in which we can see
easily that the weak value must be real, namely when the operator $O$ is
Hermitean/real, and when a narrow bunch of paths $history$'s dominates
the functional integral, i.e, what you can say 
would happen in a classical approximation, when the initial and
final states $|i>$ and $|f>$ correspond to such a classical development. 
(Of course, if essentially only one history dominates, and $O(t)$ is
the real function of the (real) variables used in the functional integral,
you really just ask for the value of $O(t)$ on the classical path dominating
and that is of course real).

In the classical approximation, one could indeed use this weak value to
extract some classical paths.

It is well known that one of the uses of the functional integral
formulation is that you can obtain the classical equations of motion
by requiring the functional derivative $\delta$ of the integrand
$\exp(\frac{i}{\hbar} S[history])$ to be zero. It even gives the classical
equation of motion in the prequantum mechanics way as the requirement
of the action being extremal.

However, if this classical use of the functional integral was the main
application, we would {\bf {not need the} $i$}!

Thus, if I say {\bf {I only care for the classical approximation}}, my only
quantum mechanics idea left over is {\bf {a functional integral of some
 sort}}, then I can just {\bf {throw away the mysterious} $i$ {if I
 do not like it}}. I would obtain a classical solution
anyway, but now I would not obtain a lot of them as with the $i$; no,
I would likely obtain only one or a {\bf {very little set of classical solutions}}
that would correspond to the truly maximal functional integrand. (Most other
solutions would only survive as saddle points, or for a very special choice
of the $<f|$ and $|i>$).

Note that the \emph{{history}} that comes out as the winner that dominates in the case,
when I leave out the $i$, still obeys the equations of motions, as if we
had the $i$, because just multiplying equations of motion with $i$ or
with any complex number ($\ne 0$) does not make any change at all. Thus, in this
exercise of leaving out the $i$, one obtains the same equations of motion, but
now of course there is only one or very few {$histories$} for which the
integrand $\exp(\frac{1}{\hbar} S[history])$ is maximal and presumably
dominates in the integral over all the other $histories$. This {$history$} is now
selected by the functional integral formalism with the $i$ left out.
We will, of course, assume that the {$history$} or {$histories$} with the very
biggest
integrand in the functional integral without the $i$ should now be the
one (or a few) realized in the world. That is to say, what really happens in the
world should be described by this maximal action solution. That is, so to speak,
the model for the {\bf {initial conditions}}, because by leaving out the $i$ the
functional integral turned into a model, for the laws of nature, {\bf
 {uniting the equations of motion with the initial conditions}}.


The reader might easily see how theories of the type with $i$ removed in
functional--integral formulation leads to a prediction of the solution to the
classical solution to be chosen, by simply noticing that the integrand
in the functional integral is (exponentially) much larger for the
$-S[hisoty] $ corresponding to a history that makes this $-S[hisory]$ larger than
the one for which it is smaller. 

\subsection{Earlier Use of Turning the Phase and Throw an $i$ Away}

The idea of making an action or Lagrangian with the phase change
like here is not quite new in the sense that you can consider Wick rotation~\cite{Wick} used in evaluating loop integrals as very similar. However, in the Wick rotation,
it is purely a mathematical calculation method, while our idea is
philosophically very different as far as we want to change a priori
physics, by taking it that the fundamental physical action has indeed been
turned in phase relative to the usual one.

In the attempts to study the holography of the Maldacena conjecture type,
it is often used to compare under the name of correlation
functions~\cite{correlation}
what we below define as ``weak values''~\cite{wv} of say the conformal
field theory
CFT and an $Ads_5\times S_5$ string theory. Here, when the question is about
testing if
two theories are equivalent, it is not so important if one puts an extra $i$
in or not, provided one does the same for both theories.

Moreover, Hawking in his theory for gravity allows himself to have
the signature of the metric changed, a modification basically like
that of our $i$ shift. When he now uses this formulation
together with his and Hartle's no-boundary postulate~\cite{noboundary}, it is
even philosophically
the same game as ours. Thus, we shall consider Hartle and Hawking as the
forerunners
of the present work.

\subsection{Could This Initial Condition Model Have Any Chance at All?}

To achieve any idea about how such an ``$i$-removed'' might predict
the chosen solution to the (classical) equations of motion
(meaning choosing intimal conditions),
we might think of a slightly simpler system, rather than the whole world with
its field
theories and gravity with black holes and other strange configurations and an
energy concept that requires a special invention/or gauge choice, to make a system that might be nice
to \mbox{think about.} 

A still very general system that we could think about, and into which we might
put a more primitive version of gravity, would be a, non-relativistic for
simplicity, particle running in a potential in say a finite dimensional
space, say dimension $N$. 
Such an particle in $N$ spatial dimensions really
could be interpreted as many $N/3$ particles,
namely by letting the $q_i$s be coordinates of the first three for
the first particle, the next three for \mbox{the second, \dots} 

For such a system, we have an action
\begin{eqnarray}
 \hbox{action } S[q] &=& \int L(q, \dot{q})dt\\
 \hbox{where } L(q,\dot{q}) &=& T(\dot{q})-V(q)\\
 \hbox{and } T(\dot{q}) &=& \sum_{i =1}^N\frac{1}{2}m_i\dot{q}_i^2\\
 \hbox{and } V(q) &=& ``\hbox{a potential function of} q''. \end{eqnarray}

Here, we have let $q$ be a set of variables
\begin{eqnarray}
 q&=& (q_1,q_2,\dots,q_{N-1},q_N) = ( \hbox{ordered set} \; q_i|i=1,2,\dots,N)
 \nonumber.
 \end{eqnarray}

We can imagine a complicated landscape with many peaks and valleys in the
general potential $V(q)$. Then, we have to think of almost a ``god''
(a ``god'' in quotation marks) that has to figure out/calculate what
classical solution will maximize this action $S[q]$.

If really we ask for such a maximal action, then the solution would be
somewhat bad, because the kinetic energy is not bounded from above.
Thus, if we seek a solution with as high a kinetic energy as possible, then
the motions of the particles would be infinite or there should be some cut off,
if it should make sense.
In any case it would mean that the description with the particle variables
we started from would not be good. In such a world, one would probably have
used some different variables.
However, instead of philosophizing here about what such divergently running
variables could mean, if anything, let us just take the opposite sign
for the action put into the functional integral \mbox{integrand exponent.}

We are allowed to think that the world without the $i$ had replaced
this $i$ by $-1$ instead of just by $1$, because the sign does {\bf {not}}
matter for the classical equations of motion. Thus, to avoid divergence
of the kinetic energy (which might be postponed to a later work), we put the
sign in the exponent in front of the action to be minus, so that the
favorite solutions (the ones with high probability) have the highest potential
energy and lowest kinetic one, as far as these can be combined.

This means we
choose the opposite sign, namely to take the functional integral
\begin{eqnarray}
 \int \exp( -\frac{1}{\hbar} S[history]) {\cal D}history\\
 \hbox{where } history &=& \hbox{an ordered function set } q :
 {\hbox{time axis}} \rightarrow {\bf R}^N.\nonumber 
\end{eqnarray}

Now, we make the ``god'' seek a solution, in which the world stands
most of the time on top of the highest peaks in the landscape potential
$V(q)$. (We assume to avoid problems an upper bound for the potential
energy $V(q)$).

However, with this sign, and with just a tiny shaking, it may happen that the world/ particle
slides down from the very peak, and now, the equations of motion will make
it run down faster and faster from the peak, like a skier without the
ability to break. (I remember a tour where I had come out together with
Paul, but we were alone, when my ski made such a tour. Luckily for me I was not
on the ski when it went down with high speed. As I was now walking in deep snow and
slowly went down myself, people began to ask about if I had insurance.
I began to fear the ski had hurt or killed somebody. However, shortly after,
I saw the ski planted up down in the snow, but alas in two pieces.
Oh dear. However, the ski was rented, and seemingly insured. The kind renter
offered me a new ski for the rest of the day, but I thanked him no, and
went on doing physics instead; it was enough for that day).

It is difficult to see that such a situation should not end in a total
catastrophe in
which the
potential $V(q)$ which now comes into the exponent with a plus sign should go
lower and lower and thus make a soon negligible contribution. Such a
solution could
of course not be the true winning dominant classical solution. What
shall the ``god'' wishing a highest possible potential energy minus kinetic one 
do?

The solution to avoid the catastrophe preventing any hope of having the
dominant $history$ in the functional integral must be to arrange a way
up to some neighboring hill-top as quickly as possible.

(Remember that we are just seeking the maximal ``-action'' solution, and that
means, that the ``god'' that shall find it, also has the power also to arrange the
future as good as ``he'' can by adjusting and fine-tuning the initial state).

\subsection{Is This Scenario a Caricature of Cosmology?}

We may optimistically interpret this to be a crude picture of what
goes on in cosmology with regard to to the very strongest events:

\begin{itemize}
\item Standing on the highest peak as long as possible could be
 identified as a slow-rolling inflation: The inflation field
 stands on the highest place in the inflaton potential. It stands
 by making the uttermost effort to keep there as accurately as it can
 be arranged, and the physicists think it stayed too long to be believable
 and call it the {\bf {slow roll problem}}, because it stayed so long.

 Finally, it could not avoid a bit of shaking, and at the end, the inflaton
 field \mbox{rolled down.}

 It is like putting a pen to stay just on the tip, just at the meta stable
 point, and then find it there for years. If nothing else would shake it,
 even a very weak quantum effect would do the job, and getting such a pen to
 stand straight up for a long time is not possible in practice.

\item Next, it should run up a neighboring hill, and it would be better to be
 equally as high so that it can stop there again for a very long time.

 The way {\bf {up}} is to let all the particles be shut away
 from each other, so that we have the gravitational potential growing, when
 particles go away from each other.

 The idea in the real world is that we can claim that the advice the
 ``god'' takes to come up on a high hill again quickly is to make a
 very strong Hubble--Lemaitre expansion from at least the end of the
 inflation. (Really, there is already a Hubble--Lemaitre expansion going on
 during
 the inflation, so it might be rather easy to achieve for ``god'' to just
 continue that; or rather, ``he'' arranged inflation in the inflation
 after his purpose also after inflation).

 Then, rather soon, most of the kinetic energy of the material/the particles
 should be converted into the potential energy from them being
 separated (think
 about a Newton gravity approximation). If this is carried out, then
 as the system approaches the next peak, the kinetic energy would be more and
 more suppressed and the system should be moving less and less.

 In the Universe as we know it in cosmology, the contributions to the energy
 density counted in the usual LFRW coordinate choice, which are expected to
 have much kinetic energy in them, should
 have been suppressed by the arrangement to find a {\bf {potential}} peak.

 Now, the galaxies, etc., have run so far away that they have run most of their
 original speed off in the sense that the Hubble--Lemaitre expansion has
 dropped
 very much compared to the original one, say at the end of inflation.

 Now, physicists believe that the Universe should continue to expand forever,
 but not terribly many years ago, one believed it could contract again some day;
 with
 possible doubts about what dark energy really is and the uncertainties
 in measuring it, we should rather say: only believing the strongest and most
 certain effects, it could well be that the universe would still slow down
 and approach a null Hubble--Lemaitre constant. That would correspond to
 stopping on the hill-top. In any case, the expansion rate is minute
 today compared to what it was.
\end{itemize}

\subsection{Seeking FLRW Formulation}

In general relativity, even the concept of energy is coordinate choice dependent, and thus, we shall here, rather than making a full general relativity
formulation, choose to consider a little subset of galaxies or just
dust particles and study their kinetic energy, say in a frame of the center of
mass of the little subset. If we take it that the scale factor $a$ in the
usual Friedman--Lemaitre--Robertson--Walker (FLRW) formulation is a true length
scale, say
the radius
of the universe, and consider a subset of the galaxies of
unit length extension (maybe the unit is Gparsec), then the time
derivative $\dot{a}$ is the typical velocity in the essentially flat space
frame for the little subset with its center of mass taken to be at rest.
Now, we may look at the FLRW equations
\begin{eqnarray}
  \left({\frac {\dot {a}}{a}}\right)^{2}+{\frac {kc^{2}}{a^{2}}}-
       {\frac {\Lambda c^{2}}{3}}&=&{\frac {\kappa c^{4}}{3}}\rho\label{FLRW1}
       \\
       2{\frac {\ddot {a}}{a}}+\left({\frac {\dot {a}}{a}}\right)^{2}
       +{\frac {kc^{2}}{a^{2}}}-\Lambda c^{2}&=&-\kappa c^{2}p.
\end{eqnarray}

 The goal of the theory (the ``god'') without the $i$ is to avoid
kinetic energy and keep the energy as potential energy as much as possible,
and that will mean concentrating on the little subset of galaxies to keep
the time derivative of the scale parameter $\dot{a}$ as small as we can
weighted with time. Thus, it is most important to keep the velocity $\dot{a}$
small in the long time intervals. 

Now, it is well known and easy to see from the FLRW Equation (\ref{FLRW1})
assuming
a dust model (matter dominance), otherwise the velocity of a galaxy makes
is not important, that the following cases can happen:
\begin{itemize}
\item {\bf {If}} the parameters, such as the $\Lambda$, are so large and positive
  that the Universe, as it seems empirically for the time being, goes into
  a $\Lambda$-dominated {\bf {DeSitter universe in the long run}}, then
  \begin{eqnarray}
    \frac{\dot{a}}{a} &\rightarrow& costant \\
    a &\rightarrow& \infty\\
    \hbox{so that } \dot{a} &\rightarrow& \infty
  \end{eqnarray}
  and our formal kinetic energy goes to infinity. Thus, this case
  would {\bf {not be chosen}} in our model.
\item {\bf {If}} on the other hand the $\Lambda$ is so negative that
  the {\bf {Universe recontracts}}  at some time, then, with the matter dominance
  ansatz
  \begin{eqnarray}
    \rho &\propto& \frac{1}{a^3}\\
    \hbox{when $a$ gets small } \dot{a} &\propto&\frac{1}{\sqrt{a}}\\
    \hbox{and thus } \dot{a} &gets& \hbox{huge}.
  \end{eqnarray}
  Thus, this {\bf {should}} also {\bf {not be chosen,}} (if one wants to minimize
  kinetic energy).
\item {\bf {If}} the Hubble--Lemaitre constant $\frac{\dot{a}}{a}$ keeps
  from both growing up and from turning down, then there is a chance that
  the velocity $\dot{a}$ could be kept small. However, the most favorable and thus
  our model prediction would be
  to have
  \begin{eqnarray}
    \dot{a} &\rightarrow& 0\\
    \hbox{even though } a&\rightarrow& \hbox{large} \hbox{ or } \infty\\
    \hbox{but then } \Lambda  &=&0.
  \end{eqnarray}
  This is the case we must have to avoid the kinetic energy in the long run.

  Thus, indeed, the {\bf {prediction is that} $\Lambda =0$.}

  If we require there to be a very small velocity $\dot{a}$ over a long
  time once
  the $a$ goes through very large values, so that the density $\rho$ has gone
  almost to zero, then, in fact, both the $\Lambda$ term and the
  space curvature term $k/a^2$ have to be zero. Thus, we see that our
  ``throwing away the $i$'' model indeed predicts both that the universe is flat,
  i.e., $k/a^2\sim 0$,
  and that the cosmological constant $\Lambda$ be ``zero''.
  
\end{itemize}

We can certainly see that to the very first approximation, namely if
we compare to energy densities in the reheating era, the cosmological constant
$\Lambda$ is indeed minute, and thus, the prediction that it should be zero is very
good. However, in today's best fit, we know that the cosmological constant is not
quite zero. However, there is still so many ways of making alternative
speculations by
replacing $\lambda$ with something else like domain walls or
quintessence running $\Lambda$, so that it is absolutely not quite excluded, and
that the cosmological constant could be avoided. Even experimental
uncertainties,
that might be needed to repair the various tensions, could also make the zero
$\Lambda$ become a possibility.

\subsection{Field versus Particle Kinetic Energy, a Little Problem?}

Even though the above description of the losing and quick recovery of
the potential energy as being very crudely arranged in our picture of
cosmology sounds by words ok, there is actually a mistake in it, in that
the rushing down the peak describing the slow roll inflation is a description
in terms of the {\bf {fields}}, whereas the description of the Hubble--Lemaitre
expansion as climbing up a gravitational potential is a description in terms
\mbox{of {\bf {particles}}.}

If we consider the fields even classically as the most fundamental description,
then we should define kinetic and potential energy in terms of fields
concerning both  {situations considered.}

To obtain an idea about how to translate between kinetic and potential
energy concepts defined for fields versus for particles, let us consider,
e.g., in~\cite{ed}, the expression for the energy momentum tensor of a
real scalar field $\phi(x)$ (Klein--Gordon equation field) 
\begin{eqnarray}
 T_{\mu\nu}&=& \phi,_{\mu}\phi,_{\nu} - \frac{1}{2}g_{\mu\nu}
 g^{\rho\tau}\phi,_{\rho}\phi,_{\tau} -\frac{1}{2}m^2\phi^2g_{\mu\nu}\\
 \hbox{so that energy density }T_{00}&=& \phi,_0\phi,_0-\frac{1}{2}
 g_{00}g^{00}\phi,_0\phi,_0 -\frac{1}{2}g_{00}\phi,_i\phi,_i-
 \frac{1}{2}m^2\phi^2g_{00}\nonumber\\
 &=& \frac{1}{2}\phi,_0\phi,_0 +\frac{1}{2}(1+2\varphi(x))\phi,_i\phi,_i+
 \frac{1}{2}(1+2\varphi(x))m^2\phi^2\nonumber\\
 &=& \frac{1}{2}\phi,_0\phi,_0+\frac{1}{2}(1+2\varphi(x))(\phi,_i\phi,_i+
 m^2\phi^2)\label{25}
 \end{eqnarray}
 where we assumed $g_{ii} = \eta_{ii} = 1 $ for the spatial
 indices, and $i=1,2,3;$ and in flat space, $g_{00}=-1$.
 However, in a Newton gravity approximate situation
 \begin{eqnarray}
 g_{00}(x)&=& -1 -\frac{2\varphi(x)}{c^2}\\
 &=& -1-2\varphi(x)
 \end{eqnarray}
 where $\varphi(x)$ is the gravitational potential.

 It is obvious that you could look naively at Equation (\ref{25}) and see the
 first term $\frac{1}{2} \phi,_0\phi,_0 $
 {is in the field
 terminology the kinetic term, while the second term
 $\frac{1}{2}(1+2\varphi(x))(\phi,_i\phi,_i+
 m^2\phi^2)$} is the potential one, because the first term
 consists of the time derivatives $\phi,_0$ of the Klein--Gordon field
 $\phi$, while the second term has only the $\phi$ field itself, without the
 {\bf {time}} derivative. There is a gradient term with spatial derivatives
 $\frac{1}{2}(1+2\varphi(x))\phi,_i\phi,_i$, but that must principally
 be counted as potential. We can see that interaction with the
 Newtonian gravitational field $\varphi(x)$ is solely in the potential
 part.

 It looks promising for the potential energy for the particle language
 being identical to a part of the field-wise potential energy, but
 I do not trust that. We must investigate/think about what really
 happens when such a Klein--Gordon field describes a particle coming
 with significant velocity and run up a slowly varying Newtonian
 gravitational potential~$\varphi(x)$. That the particle is running
 with significance 
  (but we can for simplicity still think that it is
 non-relativistic, not to be nervous about using Newton approximation)
 means that there is at first a gradient part of the energy, but that is
 counted as potential field-wise. However, as long as the particle has a
 rather well-determined momentum in spite of being localized relative to
 the very little variation in space of the gravitational potential, then
 the vibrations of the Klein--Gordon field will be like a harmonic
 oscillator and there will be necessarily equally as much potential
 and kinetic energy provided when one counts the zero for the potential energy
 at the minimum in the approximate harmonic oscillator potential. This
 means that the two terms we pointed out as
 kinetic and potential energy, respectively, will be approximately equally as
 big when integrated over the relatively large region to which the
 particle is located. This approximate equality is to be understood
 with the normalization, in that both terms would be zero if the Klein--Gordon
 field $\phi$ were zero. This means that, some time later, when the
 particle has climbed up a hill and come to a region where the Newtonian
 gravitational potential $\varphi(x)$ is higher, this normalization of
 the potential is different.

 Now of course what happens under the climbing up of the particle
 is that it slows down, and that means that the gradient term
 will be smaller when it has come higher up. Both the kinetic and potential
 term will have sunk, counted in the local normalization, meaning that they would be zero when the Klein--Gordon field is zero.

 Thus, we see that, indeed, by climbing up, the kinetic energy term in the
 field-wise sense will have fallen. 
 The potential term will also have fallen if one uses the local normalization, but using the same
 normalization it must have increased because the total energy should
 be conserved.

 Thus, after all we came to, even when strictly using the definition
 of kinetic and potential energy separation by means of the fields
 (i.e., the Klein--Gordon field $\phi$), then we find that the climbing of
the hill indeed gives a conversion of the kinetic energy to potential
 energy as the naive thinking in the particle definition of the
 the kinetic and \mbox{potential energies.}

 Thus, indeed, the making of the Hubble--Lemaitre expansion to make
 the kinetic energy into potential energy is a good idea for the ``god''
 to perform, even when we take the field definition of these
 concepts of various energies as the fundamental one.

 However, it should be noted that defining kinetic energy and potential
 energy by means of fields or by means of particles is {\bf {not}} quite the
 same:

 We calculated here, as is easily seen, the energy of the particle
 in excess of the gravitational potential to be {\bf {half potential
 and half kinetic.}}

 In the particle definition, however, the energy due to the motion of the
 particle is counted as {\bf {purely kinetic}}. The Einstein mass energy is a bit more doubtful for the particle and should presumably be counted as
 potential.

 \subsection{What to Think about Black Holes for the ``God'' ?}

 It sounds obvious that if the ``god'', in quotation marks, is so
 eager to make kinetic energy into potential energy, so that he makes/arranges the material in the Universe to Hubble--Lemaitre expand
 dramatically to achieve that, he would consider making black holes,
 primordial or later, as a kind of catastrophe, since it would undo
 ``his'' great efforts with the expansion. Thus, ``he'' should put the
 density of black holes as low as is possible for ``him'' within
 ``his'' competence of arranging details in initial conditions.

 However, even ordinary matter and dark matter of a different nature
 than black holes should be kept down in amounts, because, as I claimed above, half of the Einstein mass energy that should be for a Klein--Gordon field particle could be counted as kinetic and half as potential.

 Thus, even having ordinary or non-black-hole dark matter seems
 not to be wanted.

 One can find comfort in the idea of the determination having high
 potential energy by telling that in the first 70,000 years or so
 we had a radiation-dominated universe, so that these massive
 particles came in after a major work of climbing up the next
 hill is already over in the first approximation. 

 Thus, one could still claim that the first and strongest
 Hubble--Lemaitre expansion was performed without any problem of
 thinking about the masses of the surviving particles.
 It was, namely, radiation dominance.

 \subsection{Our Dark Matter Model from the Point of View of the
   ''God'' in Quotation Marks}
 The ``god'' in quotation marks sat and read Paul Frampton's work that dark
 matter could be black holes, and thus, at the time of the humans, there
 should be about 24 percent, after the mass/energy of the
 density, that should be dark matter. ``He'' got rather sad by reading that
 and said to himself: ``This is terrible when I now have made so much
 effort to expand the universe so as to make the
 (gravitational) potential energy so positive as possible, then to spoil
 it by
 making such a lot of black holes. That theory shall never be right!'' However,
 then, he thought a bit and added, ``Well, there
 are many proposals that are not going to be true'' This encouraged ``him'' a
 bit, but then he thought on, ``But I am also in a theory that is
 proposed, but that proposed theory is even worse than that of the dark matter being black holes, so I am very likely not true, I do not
 exist! It is very much more likely''. Then, ``he'' became very
 depressed: ``It
 is to be so much dark matter and in addition I do
 not even exist''. He now became so depressed that he almost lost all
 courage for life. He almost thought it was better not to exist at all.
 
A little later (whatever that means for a
timeless ``god''), he sat and googled quite without interest; ``he'' really was
seriously depressed. 

\begin{figure}
  \includegraphics[scale=2]{
  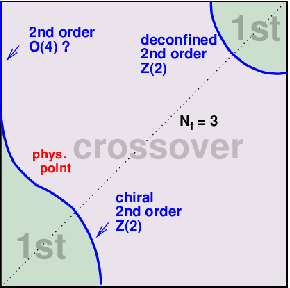}
 \caption{{``Columbia-plot'':} on the abscissa is taken to be common mass of the
 up and down quarks, and on the ordinate, the strange quark mass. By
 simulation, one has for each combination of quark masses looked for whether
 increasing the temperature leads to a phase transition of first order or
 only a so-called crossover, which means that there is no genuine phase
 transition
 as function of temperature, but possibly a rather steep variation of an
 order parameters. It is the phase transition separating curve, which
 separates
 the lower left corner, the zero quark mass point, from the middle of the
 diagram, which is the phase transition, that ``god'' wants to use to
 make the Universe have two sorts of vacua.
 The quark mass axes goes from 0 to $\infty$.}
\label{figure2}
\end{figure}

The ``god'' in quotation marks was so sad that he was googling almost at
random and among other things, ``Columbia-plot''~\cite{Columbia} (see Figure~\ref{figure2}) came about and he saw that
lattice QCD calculations led to there being a phase transition, of second
order, but anyway, as a function of the light quark masses of the vacuum.

Now he got the idea that he could put the quark masses just on the phase
transition curve. It was not too far from where he would have thought of
putting them anyway. Then, he could obtain different phases of vacuum that were
realized in Nature in different places and times, and he could hope
for some surface, a domain wall between the phases, and of course, there would
be a bit of potential difference between the two phases for the nucleons. He
got the idea of using such potential difference to catch the nucleons as in a
jail and let the skin between the phases contract itself around the caught
nucleons and squeeze them so hard that he could catch very many in a small
region. Then, he would have very heavy pearls of jails for nuclei and could
make use of that in two ways:

{\bf {First}:} The pearls would pick up electrons of course and become almost
neutral, but in
any case, relative to their size, hugely heavy and thus function as dark matter.

{\bf {Secondly}:} He could get rid of most nucleons and thus most matter, and
thus of most stars that could finally even develop into neutron stars or
even worse to black holes. It would be wonderful: he would only have some
dark matter that would not be able to clump severely into clumps where the
gravitational potential would become decreased so that potential energy
was converted into kinetic energy. He was so glad for this idea, because if
he could not get rid of the baryons totally, this was the best he could do:
to catch and keep them from interacting and losing their potential energy
into kinetic energy. 
 The clumping would be much less than for ordinary
matter that was not jailed. Thus, no serious 
clumps could be formed and decrease the potential energy of gravity, and
nothing of the bad stuff such as stars and black holes should come out of it.
He had an alternative to the dark matter being black holes, and he got rid of
the ordinary matter that had a tendency to make stars and later even black
holes. Hurrah, he became very happy.

However, suddenly, he heard some explosions, it seemed! What was that? He looked
and saw that inside his small jails for the nucleons, fusion bombs had
appeared. The nucleons had first formed helium under the high pressure inside
the pearls, and then, in an explosive way, the helium had also combined with
carbon. The latter happened so explosively that a lot of nucleons were
expelled out from his so smartly invented jails because of the high temperature from the
fusion of the helium to carbon. About one-fifth of the nucleons in the jails had
escaped due to the explosions, and he had got a lot of freely running nucleons back.
That was a disappointment for him, but at least most of the nucleons stayed
in the jails.

However, he was still much more happy now that the terrible scenario of the 24
percent of black holes had been avoided, and he contemplated: It was ME who
adjusted the quark masses so that we obtained two phases with the same energy
density, so that they could be in balance and coexist 
without the one immediately decaying to the other one. Thus, if HE had done
such a thing, should he not then exist? In spite of the little bad luck in that
the nucleons had made this fusion bomb and some of them had escaped, the
situation was still much better than with a lot of black holes, and a thread
of his very existence. Now, HE at least believed in himself; whether the
others would believe in him or not, he now \mbox{did himself.} 

\subsection{The Ratio of Dark to Ordinary Matter}
The explosion of helium fusion to carbon (or some other heavy elements, but it
would have been carbon most likely) was one of the first points studied by Colin Froggatt
and me concerning dark matter. In {\cite{Dtoo},}  we wrote in 2005: ``Before the further internal fusion process took place, the main content of
the balls was in the form of ${}^4He$ nuclei. Now the nucleons in a ${}^4He$
nucleus
have a binding energy of 7.1 MeV in normal matter in our phase, while a
typical “heavy” nucleus has a binding energy of 8.5 MeV for each nucleon. Let
us, for simplicity, assume that the ratio of these two binding energies
per nucleon is the same in the alternative phase and use the normal binding
energies in our estimate below. Thus we take the energy released by the
fusion of the helium into heavier nuclei to be \mbox{8.5~MeV ${-}$ 7.1 MeV = 1.4 MeV} per nucleon. Now we can calculate what fraction of the nucleons, counted as
a priori initially sitting in the heavy nuclei, can be released by this \mbox{1.4 MeV}
per nucleon. Since they were bound inside the nuclei by 8.5 MeV relative
to the energy they would have outside, the fraction released should be (1.4~MeV)/(8.5~MeV)~=~0.165~=~1/6. So we predict that the normal baryonic
matter should make up 1/6 of the total amount of matter, dark as well as
normal baryonic. According to astrophysical fits, giving 23\% dark matter
and 4\% normal baryonic matter relative to the critical density, the amount of
normal baryonic matter relative to the total matter is\linebreak 4\% /(
23\% + 4\%) = 4/27 = 0.15''.

Not so many dark matter models give the ratio of the dark to ordinary
to be of order unity directly; it more comes out as a miracle that the order
of magnitude is the same. 

 \section{Conclusions and Birthday Wishes}
 \label{conclusion}
 We have talked about the fact that it was very close, and that it would have been
 Paul and Nambu who would have been the known inventors of string theory
 and not myself and Susskind. (Both combinations in addition to Nambu).

 Additionally, we talked about a genuinely still not excluded replacement for the
 Standard Model, the 3-3-1 model(s) (there are possibilities for some
 variations of the model), a model in which one, in order to cancel the
 (triangle) anomalies, has to have the number of families being a
 (multiple of) the number colors in QCD. This is of course a remarkably
 good prediction. There are three families and three colors.

 {An important sign of the 3-3-1 model is that in the spectrum of leptons of the same sign one should find resonances signaling decaying gauge particles}.

 We also mentioned that if dark matter is indeed primordial
 black holes, they have to be rather heavy compared to many alternative
 pictures, but most severely, they would be so heavy that there would
 be so long
 between them hitting earth that the DAMA-LIBRA experiment, which in spite
 of being in contradiction seemingly with even very similar experiments
 as the Anais experiment, with very high statistics for having seen dark
 matter in the underground, should not see so much as they saw.

 At the end, I sneaked in my own crazy theory of the last weeks of
 little ``god'' in quotation marks seeking to govern the world
 so as to put most of the energy as {\bf {potential energy}}, and that
 this could be interpreted to mean that just after the ``reheating''
 time, when the inflation in which all energy was indeed potential
 ended, it was organized that a huge expansion quickly
 could convert the kinetic energy back to potential energy.

 In fact, we argued that such a ``god'' disfavoring kinetic energy
 would make the cosmological constant $\Lambda$ and curvature term
 in the FLRW--cosmological equations \mbox{preferably zero.} 
 
 If there really was such an organization that liked to arrange the
 kinetic
 energy to quickly come back to being potential, e.g., by Hubble--Lemaitre expansion,
 then of course the production of black holes in which matter falls into the
 black hole in a strong gravitational potential and thus sees its
 potential energy converted into kinetic energy, would be seen as a bad
 thing to do.
 Thus, this ``god'' would only accept black holes to the extent it would be
 almost unavoidable.
 \subsection*{Congratulation}
 Let me first give thanks for the very nice times we have had together and
 for the many discussions, etc.
 Then, the best wishes {for}  
 that the ``god'', if ``he'' exists, has
 planned a great
 future for Paul, and if he does not exist, that Paul may have a very
 lucky and successful future anyway, in the latter case, even with dark
 matter being primordial black holes. 
  Good luck with the birthday! 
  
  \vspace{6pt}

Funding: {This research received no external funding.}



Conflictsofinterest:{
  {The authors declare no conflicts of interest.}}

\vspace{6pt}



%


\end{document}